# Frequency and Spatial domain based Saliency for Pigmented Skin Lesion Segmentation

Zanobya N. Khan
Department of CS & IT, SUIT Peshawar, Pakistan

**Abstract**

Skin lesion segmentation can be rather a challenging task owing to the presence of artifacts, low contrast between lesion and boundary, color variegation, fuzzy skin lesion borders and heterogeneous background in dermoscopy images. In this paper, we propose a simple yet effective saliency-based approach derived in the frequency and spatial domain to detect pigmented skin lesion. Two color models are utilized for the construction of these maps. We suggest a different metric for each color model to design map in the spatial domain via color features. The map in the frequency domain is generated from aggregated images. We adopt a separate fusion scheme to combine salient features in their respective domains. Finally, two-phase saliency integration scheme is devised to combine these maps using pixelwise multiplication. Performance of the proposed method is assessed on $PH^2$ and ISIC 2016 datasets. The outcome of the experiments suggests that the proposed scheme generate better segmentation result as compared to state-of-the-art methods.

## 1. Introduction

Over the past few years, a rapid increase in the melanoma and non-melanoma skin cancer cases has been observed all over the world. According to the article published in [1], Pakistan is among those countries where skin cancer is on the rise especially among women and young girls. The main cause being the use of sub-standard whitening creams that contain toxic materials including vitamin E, mercury and potent steroids. The heuristic metrics developed by the dermatologists, such as ABCD rule [2], Menzies method [2], 7-point checklist [3] and CASH algorithm [4] for the diagnosis of skin cancer, are prone to clinical subjectivity and reproducibility. The need for usage of computer-assisted diagnostic (CAD) systems has increased for proper interpretation of dermoscopy images. Automatic segmentation of skin lesion from its surrounding is a challenging task as the lesion color, size, skin structure, and shape varies considerably for different patients. Complex background regions and artefacts such as hairs, color charts, and black frames make the segmentation process even more challenging.

Saliency-based techniques are primarily used as a pre-processing step to enhance the salient region. Saliency-based models have been extensively used in many computer vision applications such as object detection and content-based image retrieval systems. Saliency detection methods can be categorized as spatial domain and frequency domain. The spatial domain based methods focus primarily on the local information present in the image and can detect the details of the salient region. The frequency domain based methods, on the other hand, explore the global information of the image and can identify the main object [5]. Ahn et al.[6], utilized background information in the construction of the saliency model to detect pigmented skin lesion. However, dermoscopy images with color calibrated charts and dark regions in the dataset were manually removed and replaced with natural skin pixels before beginning the pre-processing stage. Fan et al. [7] made use of color contrast and brightness information only and proposed an improved Otsu's thresholding method for segmentation of skin lesions. Nevertheless, images with low contrast and color charts were detected incorrectly. These saliency models were designed in the spatial domain.

As opposed to saliency models of [6] and [7], in this paper we combine spatial domain and frequency domain based algorithms to detect salient region in dermoscopy images. The color contrast map is designed in the spatial domain using color features extracted from skin lesion images in COC (which corresponds to human visual system) and CIE Lab color spaces. In frequency domain, Fast Fourier Transform (FFT) is first applied to the aggregated images attained by fusing color channels of COC and CIE Lab color models. The map is obtained by integrating filtered amplitude spectrum with the original phase spectrum. The next phase involves the integration of these candidate maps to obtain the detection map. Most of the existing studies used a unified fusion scheme to integrate multiple maps. Xu et al. [8] indicate that when candidate saliency model misjudges a region on an image, the region will also be misjudged on the integrated image. In this paper, we suggest a different fusion scheme for candidate maps derived in spatial domain and frequency domain according to their features. The candidate maps are then integrated using a pixelwise multiplication to develop the final saliency map. The main contributions of this article are:

1. A different metric is proposed for each color model to generate map in spatial domain by extracting color features from input images.
2. We adopt frequency domain based map to detect lesion in dermoscopy images in contrast to previous saliency based models proposed for detection of skin lesions. The map is created using the amplitude and phase spectra.
3. An individual fusion scheme is suggested for candidate saliency maps to improve detection accuracy. The candidate maps are then integrated using a pixelwise multiplication to formulate the final saliency map.

The remaining paper is organized as follows. In Section 2, the pre-processing techniques are discussed briefly. Section 3 reviews the selection of appropriate color space. Section 4 describes the generation of map in spatial domain and frequency domain. The computation of final saliency map and its binarization is detailed in Section 6. Results and conclusion are discussed in Section 7 and 8 respectively.

## 2. Related Work

In the CAD systems developed for skin lesion, the most important and challenging task is the extraction of lesion from its surrounding region. Two of the most commonly used methods are thresholding and region growing. In [9], Silveira et al. suggested an adaptive thresholding techniques to segment lesion. In this method, the color of every pixel is compared with a threshold and if the pixel is darker than threshold, it is classified as lesion. Yuksel et al. [10] detect the lesion by adopting histogram thresholding technique based on type-2 fuzzy logic. The thresholding methods fail if there is no distinct border between lesion and the background. Snakes or active contours method [11] can be used as an alternate to thresholding to identify the shape of the suspicious lesion in dermoscopy images. Nevertheless, the presence of hair can affect the performance of snake and active contours model.

Another popular technique used for the segmentation of lesion in dermoscopy images are region based methods such as region growing [12], fuzzy based split and merge operations and Mumford-Shah algorithm [13]. In region growing algorithm, neighboring pixels or sub-regions with similar attributes color, texture or grey level are grouped into larger homogeneous region. J. Maeda et al. [14] introduced fuzzy based split and merge segmentation algorithm. In this method texture and color texture features are first extracted from input image followed by a split and merge technique performed in four steps including simple splitting, local merging, global merging and boundary refinement. The similarity between adjacent regions is determined using fuzzy based homogeneity measure. Ganzeli et al. [15], Castillejos et al. [16], and Celebi et al. [17] utilized statistical regional merging (SRM) algorithm for the detection of edges in lesion images. In SRM, the input image is reconstructed based on statistical color distribution of lesion regions. However, these methods often result in over segmentation due to color variegation and skin texture variation. Flores and Scharcanski [18] utilized a dictionary-based approach to describe compact representation of image patches. The authors suggested an unsupervised version of the information-theoretic dictionary learning technique and skin lesions were segmented using normalized graph cuts.

In recent years, deep learning methods especially convolutional neural networks have been adopted to segment skin lesion in dermoscopy images. Bi et al. [19] suggested a multi-stage method by combining the outputs from multiple cascaded fully convolutional neural network to achieve lesion segmentation. In [20], a deep residual network with more than 50 layers was introduced for automatic segmentation of skin lesion, in which various residual blocks [21] were stacked together to improve the representative capability of their model. In [22], CNNs are combined with sparse coding and SVMs to provide a melanoma diagnosis.

## 3. Proposed Work

In this section the proposed framework for the segmentation of pigmented skin lesion is discussed. After preprocessing, saliency map derived in spatial and frequency domains is used to delineate the pigmented skin lesion from its surrounding and then Otsu thresholding is applied followed by post-processing to obtain the segmented lesion object. The steps involved in the proposed framework are given in Fig 1.

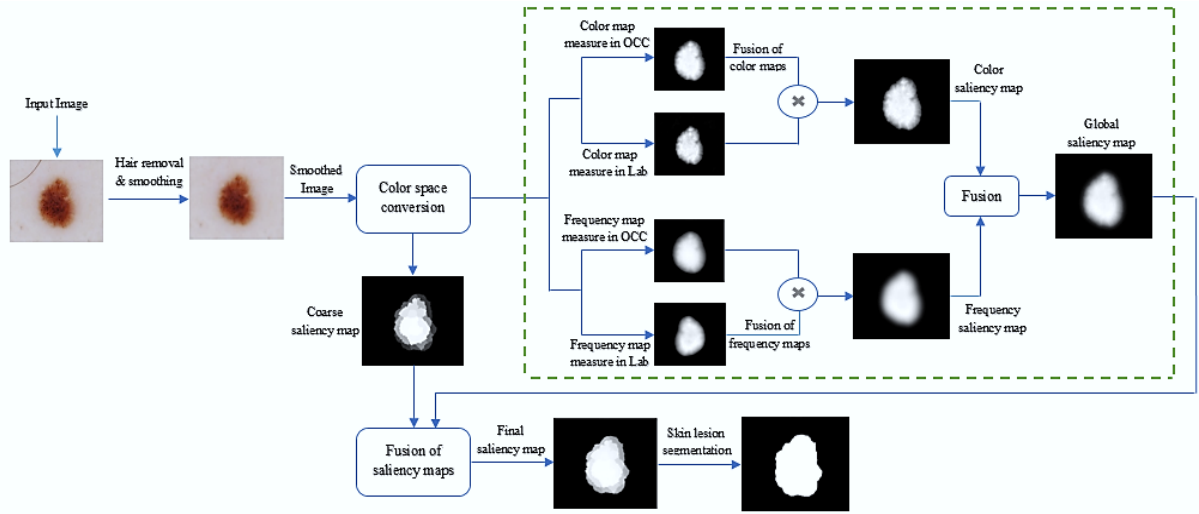

**Fig 1.** Architecture of the proposed segmentation framework

### 3.1 Hair Removal and Smoothing
The image pre-processing is significantly important for the effective detection and analysis of skin lesion in dermoscopy images as it considerably improves the overall performance of the system. The presence of visual artefacts, usually skin hair can affect the accuracy of the segmentation algorithm. In order to minimize the presence of hair in the images, the work of Lee et al. [23] is adopted. Initially, this algorithm applies a generalized gray-scale morphological closing operation separately to the three color channels to locate hairs. The use of structuring elements at three different orientation successfully detected the thick dark hair pixels. The detected hair pixels are replaced by the corresponding non-hair pixels values in the original image and the resultant image is then smoothed.

To further improve the quality of input image $I_{img}$, the influence of reflection, non-uniform illumination and variegation also need to be minimized. For this purpose, fast guided filter – an edge preserving smoothing operator [24] is applied instead of the median filter (as it is less effective against the presence of various artefacts in images). The input RGB image is used as a guidance image for the computation of the guided filter to obtain the filtered output. The image is smoothed out from these noisy factors whilst preserving borders between skin lesion and normal skin. The smoothed imaged $I_{smt}$ is shown in Fig 2.

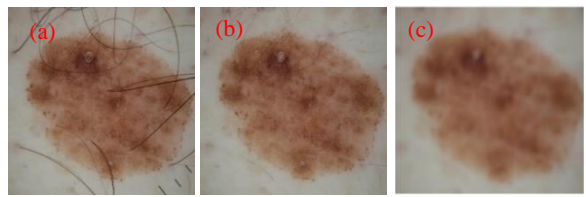

**Fig 2.** (a) Original image (b) hair removed (b) **smoothed using fast guided filter**

### 3.2 Color Space Conversion
In the analysis of dermoscopy images, color information is considered to be the most important attribute in distinguishing lesion from its background region. Most of the proposed techniques in the literature for lesion detection have used uniform and non-uniform color spaces [25]. And some of them used a combination of both uniform and non-uniform color models [26]. Since the RGB color space presents no perception uniformity and there also exists a high correlation among the three color channels [26], therefore, more suitable color spaces have to be chosen. The selection of a color space will have an impact on the accuracy of the segmentation algorithm. Therefore, two color spaces are adopted in the proposed work as one color space may not always works well [27].

The smoothed image $I_{smt}$ is converted to COC space which corresponds to the human visual system [28] and also to perceptually uniform color space such as *CIE* $La^*b^*$ [29]. Both these color spaces have produced a promising result in interpreting the color properties of images to detect salient regions. In *CIE Lab* color space, *L* represents the lightness whereas $a^*$ and $b^*$ represent two chromatic channels. From $r$, $g$ and $b$ channels of the input image, four broadly-tuned color channels R, G, B and Y [30] [31] are devised.

$$\begin{cases} R = r - (g+b)/2 \\ G = g - (r+b)/2 \\ B = b - (r+)/2 \end{cases} \quad (1)$$

$$Y = r + g - (2 \times |r-g|) + b \quad (2)$$

Then the two opponent color pairs, red-green (RG) and blue-yellow (BY), which correspond to double opponency cells in primary visual cortex of the human brain, are formulated using (1) and (2). And the intensity channel $I$ is obtained as:

$$RG = R - G \text{ and } BY = B - Y \quad (3)$$

$$I = (R + G + B)/3 \quad (4)$$

### 3.3. Spatial Domain based Map

Most of the prior studies have used more than one color space to detect salient region and achieved satisfactory result. Nevertheless, with different color spaces, they used the same metric to obtain salient features [27][32]. In this paper, we suggest that if two color models are used then two different metrics shall be adopted for saliency generation as well. The color contrast map in COC is derived from $I_{smt}$ using $L_1$ norm as opposed to Euclidean distance [33]. The use of $L_1$ norm showed more robustness towards the presence of outliers in skin lesion images and is also computationally efficient [34][35]. To further enhance the salient region, square root of L1 norm is taken.

$$Col(x,y)_{coc} = \sqrt{|(RG, BY, I) - (RG_\mu, BY_\mu, I_\mu)|} \quad (5)$$

$RG_\mu$, $BY\mu$, and $I_\mu$ represent the mean values of color channels and $|.|$ represent $L_1$ norm.

Since perceptual differences in Lab color space are approximately Euclidean [36], we exploited normalized Euclidean distance to construct contrast map in Lab color space.

$$Col(x,y)_{lab} = \sqrt{\frac{\left((L,a,b) - (L_\mu, a_\mu, b_\mu)\right)^2}{(L_\sigma, a_\sigma, b_\sigma)}} \quad (6)$$

The normalized Euclidean distance proved to be more effective in enhancing pixel values of pigmented lesion and suppressing the background as evident from visual outcome depicted in Fig. 3.

The final color contrast map $Col^{map}$ in spatial domain is obtained using pixel-wise multiplication and convolved with edge preserving guided filter. Where θ ranges from 0° to 120°. The salient region detected in spatial domain is shown in Fig. 4.

$$Col^{map} = Col(x,y)_{coc} \otimes (Col(x,y)_{lab})^{\tan \theta} \quad (7)$$

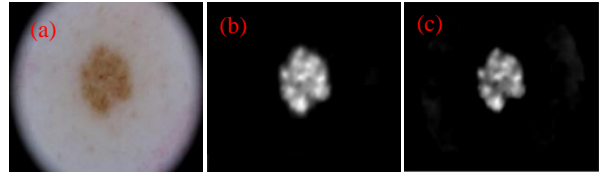

**Fig 3**. Contrast map in Lab color space using (b) standardized Euclidean distance (c) Euclidean distance

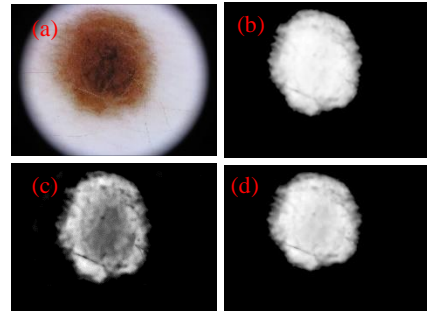

**Fig 4.** (a) Original image (b) Contrast map in COC space (c) Contrast map in Lab color space (d) Saliency map in spatial domain

### 3.4 Coarse Saliency Map in Spatial Domain

The work of Zhao et al. [37] is adopted to further improve the performance of proposed detection scheme. The patches from image background were used to produce a group of spaces of background-based distribution in order to formulate the coarse saliency map. Primarily, two color models were used namely COC and CIE Lab for feature extraction. Based on color information, patches were categorized into four different groups and represented in terms of eigenvectors and eigenvalues. As opposed to Mahalanobis distance (a data-driven metric [38]) used by authors, we resort to Euclidean distance to determine whether a patch belongs to the background or salient region. The coarse saliency map represented as $C^{map}$ is then obtained by taking the weighted average of four distance maps denoted as $\{S_p^{th}\}_{p=1}^4$. For mathematical details of this method, readers can refer to the original work [37]. Fig. 5 depicts the result of a coarse saliency map.

$$C^{map} = \sum_{p=1}^{4} w_{cs} \times S_p^{th} \quad (8)$$

### 3.5 Frequency Domain based Map

In frequency domain, two color models are employed for saliency estimation as well. However, to reduce computational cost and detect lesions accurately, frequency maps are derived from aggregated images. In COC space, a weighted sum mechanism is adopted.

$$A(x,y)_{opp} = w_1 RG + w_2 BY + w_3 I \quad (9)$$

$w_1$, $w_2$, and $w_3$ are the corresponding weights with values in the range of 0 -1. In CIE Lab color space, chromatic channels a* and b* are fused to produce weighted aggregated image $A_{lab}$. The lightness component L is not utilized due to non-uniform illumination and vignetting.

$$A(x,y)_{lab} = w_1 a + w_2 b \quad (10)$$

Values of $w_1$ and $w_2$ ranges between 0 and 1. Initially, FFT is applied on the aggregated images.

$$\begin{cases} \mathcal{F}(x,y)_{coc} = \mathfrak{F}(A(x,y)_{opp}) \\ \mathcal{F}(x,y)_{lab} = \mathfrak{F}(A(x,y)_{lab}) \end{cases} \quad (11)$$

The log amplitude and phase spectrums of the Fourier transformed images are derived as:

$$\begin{cases} \mathcal{A}(x,y)_{coc} = \log(\mathcal{F}(x,y)_{coc}) \\ \wp(x,y)_{coc} = \phi(\mathcal{F}(x,y)_{coc}) \end{cases} \quad (12)$$

$$\begin{cases} \mathcal{A}(x,y)_{lab} = \log(\mathcal{F}(x,y)_{lab}) \\ \wp(x,y)_{lab} = \phi(\mathcal{F}(x,y)_{lab}) \end{cases} \quad (13)$$

The amplitude spectrums $A(x,y)_{coc}$ and $A(x,y)_{lab}$ are convolved with the log Gabor filter $g\eth$ to suppress non-salient regions.

$$\begin{cases} \mathcal{A}(x,y)_{coc} = g\eth * \mathcal{A}(x,y)_{coc} \\ \mathcal{A}(x,y)_{lab} = g\eth * \mathcal{A}(x,y)_{lab} \end{cases} \quad (14)$$

The saliency map is generated by taking inverse Fourier transform of the new amplitude spectrum and original phase spectrum.

$$\begin{cases} Fmap(x,y)_{coc} = \sqrt{\|\mathfrak{F}^{-1}[e^{(\mathcal{A}(x,y)_{coc} + \wp(x,y)_{coc})}]\|} \\ Fmap(x,y)_{lab} = \sqrt{\|\mathfrak{F}^{-1}[e^{(\mathcal{A}(x,y)_{lab} + \wp(x,y)_{lab})}]\|} \end{cases} \quad (15)$$

The maps computed in CIE *Lab* and COC color spaces are integrated to develop resultant frequency map $Feq^{map}$ using weighted average and is convolved with a Gaussian filter $\mathfrak{G}$.

$$Feq^{map} = \mathfrak{G} \times (Fmap(x,y)_{coc} \oplus Fmap(x,y)_{lab})/2 \quad (16)$$

Where $\oplus$ denotes addition. The final result of frequency domain based map is given in Fig. 6. Further, we used different color models in frequency domain because some salient regions may go undetected in one color space can be detected in another color space as depicted in Fig. 7.

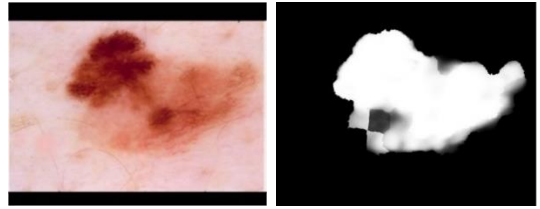

**Fig 5**. Skin lesion detection with coarse saliency map

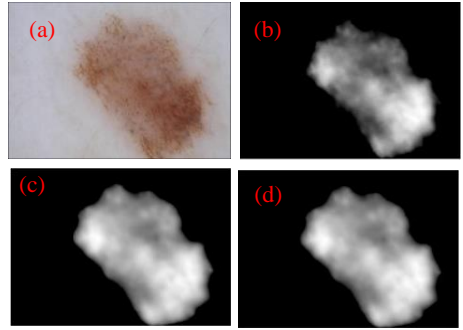

**Fig 6.** (a) Original image (b) Frequency map in COC color space (c) Frequency map in Lab color space (d) Fusion of (b) and (c) resulted in final frequency domain based map

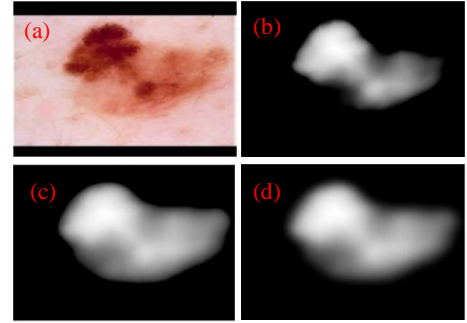

**Fig 7**. (a) Original image (b) Lesion detected in frequency domain with OPP model (c) Full lesion detected with Lab color model (d) Map after fusing (b) and (c)

### 3.6 Computation of Final Saliency Map
We propose two-phase saliency integration scheme for effective detection of skin lesion. The $Col^{map}$ and $Feq^{map}$ maps are integrated using a weighted sum to obtain initial map.

$$I^{map} = w_c \times Col^{map} \oplus w_q \times Feq^{map} \quad (17)$$

$w_c$ and $w_q$ are computed by taking reciprocal of entropy value E of the two maps.

$$w_c = 1/E(Col^{map}) \; \& \; w_q = 1/E(Feq^{map}) \quad (18)$$

In the second phase, the coarse saliency map $C^{map}$ is then fused with the initial map $I^{map}$ to produce the final saliency map using pixelwise multiplication.

$$S^{Fmap} = (w_m \times I^{map}) \otimes (w_n \times C^{map}) \quad (19)$$

The weights $w_m$ and $w_n$ are computed as:

$$\begin{cases} w_m = -\sum_{i=1}^{n} p_i(I^{map}) \log_2 p_i(I^{map}) \\ w_n = -\sum_{i=1}^{n} p_i(C^{map}) \log_2 p_i(C^{map}) \end{cases} \quad (20)$$

The normalization procedure is applied on $S^{Fmap}$ to range maps between 0 and 1.

$$\mathfrak{N}^{Fmap} = \frac{S^{Fmap} - \min(S^{Fmap})}{\max(S^{Fmap}) - \min(S^{Fmap})} \quad (21)$$

The final saliency map $\mathfrak{N}^{Fmap}$ is transformed into a binary mask image through Otsu's thresholding method [40]. To fill-in holes and remove any remaining unwanted pixels, morphological operations of opening and closing are applied to the binary mask image. The result of the proposed detection scheme is depicted in Fig. 8.

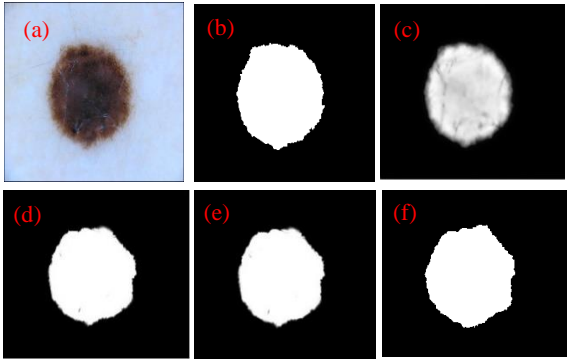

**Fig 8.** Two phase saliency integration (a) Original image (b) Ground truth (c) Initial map (d) Coarse map (e) Final map (f) binary mask of detected region

## 4 Results and Discussion
### 4.1 Datasets
To evaluate the performance of the proposed segmentation framework, experiments are conducted on two publicly available datasets: PH[2] [41] and ISIC 2016 [42]. The PH[2] dataset comprises of 200 images and ISBI 2016 contains 900 dermoscopy images. The ground truths are available for the two datasets provided by the expert dermatologists.

### 4.2 Evaluation Criteria
In order to assess the performance of the proposed method, four different statistical measures are adopted consisting of sensitivity, specificity, accuracy and dice similarity coefficient. These metrics are mathematically expressed as:

$$\begin{cases} Sensitivity = \frac{TP}{(TP+FN)} \\ Specificity = \frac{TN}{(TN+FP)} \\ DSC = \frac{2TP}{(2TP+FP+FN)} \\ Accuracy = \frac{TP+TN}{(TP+FP+FN+TN)} \times 100 \end{cases} \quad (22)$$

### 4.3 Evaluation of PH[2] Dataset
We compare the proposed saliency-based segmentation scheme with six state-of-the-art approaches: Ahn et al. [6], Fan et al [7], Rastgoo et al. [43], Bi et al. [44], Barata et al. [26] and Zhao et al. [37]. The comparison results are shown in Table I. The results show that the proposed scheme outperforms the other six methods attaining higher values for accuracy and sensitivity metrics. Also, there is no significant difference in the DSC score of Zhao et al. [37] and Ahn et al. [6] approaches because background pixels information was utilized in the formulation of their respective saliency models. The overall result demonstrates the superior performance of the proposed saliency-based segmentation framework as compared to other methods on the PH[2] dataset.

### 4.4 Evaluation of ISIC 2016 Dataset
To further illustrate the performance of the proposed method, we compare the results of the experiments conducted on ISIC 2016 with Ahn et al. [6], Fan et al. [7], Bi et al. [44] and Zhao et al. [37] approaches. The presence of artefacts, low contrast between lesion and border, and non-uniform vignetting make the ISIC 2016 a more challenging dataset. Therefore, the state-of-the-art approaches produced comparatively low segmentation results as compared to those for the PH[2] presented in Table II. Furthermore, in [6] authors replaced the color calibrated charts with background pixels manually prior to salient region detection. However in the proposed work images with color calibrated charts as illustrated in Fig. 9, we used pixelwise multiplication in (17) as opposed to a plus operator.

The results in Table II shows that our method performed relatively better against other methods used in this paper for performance evaluation. Moreover, as compared to all methods, the proposed scheme achieves a higher sensitivity, accuracy and DSC score. Zhao et al. [37] approach attained the highest specificity value (99.98%). However, low sensitivity value indicates under-segmentation because those dermoscopy images where lesion touches the boundary were falsely identified as pixels belonging to the background region.

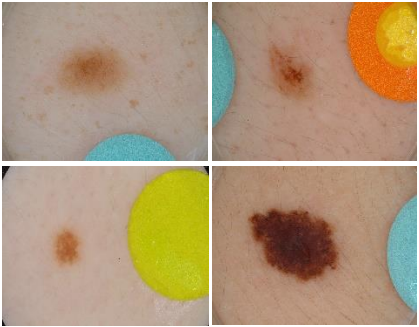

**Fig 9**. Dermoscopy images with color charts

**Conclusion**

In this article, a frequency and spatial domain based saliency algorithm are proposed for skin lesions detection. The map in the spatial domain is derived in CIE Lab and COC space. We suggested a different metric for each color space to acquire the desired results. On the other hand, frequency domain map is generated from aggregated images acquired by fusing channels in Lab and COC space. The amplitude spectrums convolved with the log Gabor filter to suppress non-salient regions are then combined with phase spectrum to devise the frequency map. A separate fusion strategy is adopted to integrate candidate maps derived in spatial and frequency domains. In order to achieve improved segmentation outcome, a two-phase saliency fusion scheme is proposed to integrate all the candidate maps. The experimental outcome based on two publicly available datasets demonstrates the effectiveness of the proposed segmentation framework

**Table 1**. Segmentation results on PH2 dataset for different methods

| Method | Sensitivity | Specificity | Accuracy | DSC |
|---|---|---|---|---|
| Ahn et al. [6] | - | - | - | 91.5 |
| Fan et al. [7] | 87.03 | - | - | 89.39 |
| Rastgoo et al. [42] | 94.00 | 92.00 | - | - |
| Bi et al. [41] | 81.57 | 88.75 | 79.87 | 95.57 |
| Barata et al. [24] | 90.40 | 97.00 | 92.80 | 90.00 |
| Zhao et al. [36] | 84.92 | 99.98 | 95.02 | 91.85 |
| Proposed | 96.83 | 94.43 | 95.23 | 93.49 |

**Table 2.** Segmentation result on ISCI 2016 dataset for different methods

| Method | Sensitivity | Specificity | Accuracy | DSC |
|---|---|---|---|---|
| Ahn et al. [6] | - | - | - | 83.90 |
| Fan et al. [7] | 74.70 | - | - | 81.83 |
| Bi et al. [41] | 78.30 | 91.31 | 85.68 | 75.88 |
| Zhao et al. [36] | 45.12 | 99.93 | 77.10 | 62.14 |
| Proposed | 79.85 | 95.15 | 88.78 | 85.57 |